\documentclass[twocolumn]{autart}    

\usepackage{graphicx}          

\newcommand{\ARAW}{A_{\rm{raw}}}
\newcommand{\APV}{A_{\rm{PV}}}
\newcommand{\qsq}{Q^{2}}
\newcommand{\gevc}{\,\mbox{GeV}^2}
\newcommand{\ges}{G^s_E}
\newcommand{\gms}{G^s_M}

\begin{document}

\begin{frontmatter}

\title{Constraints on the Nucleon Strange Form Factors at $Q^2\sim 0.1\,\mbox{GeV}^2$} 

\collaboration{The HAPPEX Collaboration}
\author[calstate]{K.~A.~Aniol},
\author[wm]{D.~S.~Armstrong},
\author[wm]{T.~Averett},
\author[syr]{H.~Benaoum},
\author[pascal]{P.~Y.~Bertin},
\author[saclay]{E.~Burtin},
\author[umass]{J.~Cahoon},
\author[uva]{G.~D.~Cates},
\author[umd]{C.~C.~Chang},
\author[jlab]{Y.-C.~Chao},
\author[jlab]{J.-P.~Chen},
\author[temple]{Seonho~Choi},
\author[jlab]{E.~Chudakov},
\author[uva]{B.~Craver},
\author[infn]{F.~Cusanno},
\author[smith]{P.~Decowski},
\author[odu]{D.~Deepa},
\author[pascal]{C.~Ferdi},
\author[jlab]{R.~J.~Feuerbach},
\author[wm]{J.~M.~Finn},
\author[infn]{S.~Frullani},
\author[umass]{K.~Fuoti},
\author[infn]{F.~Garibaldi},
\author[rutgers,jlab]{R.~Gilman},
\author[kharkov]{A.~Glamazdin},
\author[kharkov]{V.~Gorbenko},
\author[jlab]{J.~M.~Grames},
\author[jlab]{J.~Hansknecht},
\author[jlab]{D.~W.~Higinbotham},
\author[syr]{R.~Holmes},
\author[wm]{T.~Holmstrom}, 
\author[chicago]{T.~B.~Humensky},
\author[odu]{H.~Ibrahim},
\author[jlab]{C.~W.~de~Jager},
\author[rutgers]{X.~Jiang},
\author[umass]{L.~J.~Kaufman},
\author[wm]{A.~Kelleher},
\author[kentucky]{A.~Kolarkar},
\author[mit]{S.~Kowalski},
\author[umass]{K.~S.~Kumar},
\author[smith]{D.~Lambert},
\author[umass]{P.~LaViolette},
\author[jlab]{J.~LeRose},
\author[saclay]{D.~Lhuillier},
\author[uva]{N.~Liyanage},
\author[calstate]{D.~J.~Margaziotis},
\author[grenoble]{M.~Mazouz},
\author[rutgers]{K.~McCormick},
\author[jlab]{D.~G.~Meekins},
\author[temple]{Z.-E.~Meziani},
\author[jlab]{R.~Michaels},
\author[wm]{B.~Moffit},
\author[mit]{P.~Monaghan},
\author[saclay]{C.~Munoz-Camacho},
\author[jlab]{S.~Nanda},
\author[uva,petersburg]{V.~Nelyubin},
\author[saclay]{D.~Neyret},
\author[umass]{K.~D.~Paschke\thanksref{coraddress}},
\author[jlab]{M.~Poelker},
\author[kharkov]{R.~Pomatsalyuk},
\author[mit]{Y.~Qiang},
\author[jlab]{B.~Reitz},
\author[jlab]{J.~Roche},
\author[jlab]{A.~Saha},
\author[uva]{J.~Singh},
\author[uva]{R.~Snyder},
\author[syr]{P.~A.~Souder},
\author[jlab]{M.~Stutzman},
\author[kent]{R.~Subedi},
\author[mit]{R.~Suleiman}, 
\author[wm]{V.~Sulkosky},
\author[uva]{W.~A.~Tobias},
\author[infn]{G.~M.~Urciuoli},
\author[saclay]{A.~Vacheret},
\author[grenoble]{E.~Voutier},
\author[uva]{K.~Wang},
\author[harvard]{R.~Wilson},
\author[jlab]{B.~Wojtsekhowski},
\author[argonne]{X.~Zheng}

\address[calstate]{California State University, Los Angeles, Los Angeles, California 90032, USA}
\address[wm]{College of William and Mary, Williamsburg, Virginia 23187, USA}
\address[syr]{Syracuse University, Syracuse, New York 13244, USA}
\address[pascal]{Universit\'{e} Blaise Pascal/CNRS-IN2P3, F-63177 Aubi\`ere, France }
\address[saclay]{CEA Saclay, DAPNIA/SPhN, F-91191 Gif-sur-Yvette, France }
\address[umass]{University of Massachusetts Amherst, Amherst, Massachusetts 01003, USA}
\address[uva]{University of Virginia, Charlottesville, Virginia 22904, USA}
\address[umd]{University of Maryland, College Park, Maryland 20742, USA} 
\address[jlab]{Thomas Jefferson National Accelerator Facility, Newport News, Virginia 23606, USA}
\address[temple]{Temple University, Philadelphia, Pennsylvania 19122, USA}
\address[infn]{INFN, Sezione Sanit\`a, 00161 Roma, Italy} 
\address[smith]{Smith College, Northampton, Massachusetts 01063, USA}
\address[odu]{Old Dominion University, Norfolk, Virginia 23508, USA}
\address[rutgers]{Rutgers, The State University of New Jersey, Piscataway, New Jersey 08855, USA}
\address[kharkov]{Kharkov Institute of Physics and Technology, Kharkov 310108, Ukraine}
\address[chicago]{University of Chicago, Chicago, Illinois 60637, USA}
\address[kentucky]{University of Kentucky, Lexington, Kentucky 40506, USA}
\address[mit]{Massachusetts Institute of Technology, Cambridge, Massachusetts 02139, USA} 
\address[grenoble]{Laboratoire de Physique Subatomique et de Cosmologie, 38026 Grenoble, France}
\address[petersburg]{St.Petersburg Nuclear Physics Institute of Russian Academy of Science, Gatchina, 188350, Russia}
\address[kent]{Kent State University, Kent, Ohio 44242, USA}
\address[harvard]{Harvard University, Cambridge, Massachusetts 02138, USA}
\address[argonne]{Argonne National Laboratory, Argonne, Illinois, 60439, USA}
\thanks[coraddress]{Corresponding author. Email: paschke@jlab.org}




\begin{abstract}
We report the most precise measurement to date of a parity-violating 
asymmetry in elastic electron-proton scattering. The measurement was 
carried out with a beam energy of 3.03 GeV and a scattering angle 
$\langle\theta_{lab}\rangle = 6.0^{\circ}$, with the result 
$\APV =  (-1.14 \pm 0.24 \,\,\mbox{(stat)} \,\pm 0.06 \,\,\mbox{(syst)})
\times 10^{-6}$.
From this we extract, at $Q^2 = 0.099\,\mbox{GeV}^2$, 
the strange form factor combination
$G^{s}_{E} + 0.080 \, G^{s}_{M} = 0.030 \, \pm \, 0.025 \, 
\mbox{(stat)} \, \pm \, 0.006 \, \mbox{(syst)} \, \pm \, 0.012 \, 
\mbox{(FF)}$
where the first two errors are experimental
and the last error is due to the uncertainty in the neutron electromagnetic 
form factor. 
This result significantly improves current knowledge of $G_E^s$ and
$G_M^s$ at $Q^2\sim 0.1\,\mbox{GeV}^2$. A consistent picture emerges 
when several measurements at about the same $Q^2$ value are combined: 
$G_E^s$ is consistent with zero while
$G_M^s$ prefers positive values
though $G_E^s=G_M^s=0$ is compatible with the data at \mbox{95\%\ C.L.}
\end{abstract}

\end{frontmatter}

The nucleon is a bound state of three valence quarks, but
a rich structure is evident when it is probed over a wide range
of length scales in scattering experiments.
In one class of measurements, 
elastic lepton-nucleon electromagnetic scattering is used to 
measure electric and magnetic form factors, which are functions of 
the 4-momentum transfer $Q^2$ and carry information on the nucleon 
charge and magnetization distributions.

A ``sea'' of virtual quark-antiquark pairs of the three light (up, down and
strange) flavors and 
gluons surrounds each valence quark. One way to probe
the sea is to investigate whether strange quarks contribute to the static
properties of the nucleon. Establishing a 
nontrivial role for the sea would provide new insight into non-perturbative
dynamics of the strong interactions.

Weak neutral current (WNC) elastic scattering, 
mediated by the Z$^0$ boson, measures form factors that are sensitive to a 
different linear combination of the three light quark 
distributions. When combined with proton and neutron electromagnetic 
form factor data and assuming isospin symmetry, the strange electric 
and magnetic form factors $G_E^s$ and $G_M^s$ can be isolated, thus 
accessing the nucleon's strange quark charge and magnetization 
distributions~\cite{aneesh}. 

Parity-violating
electron scattering is a particularly clean experimental technique to extract
the WNC amplitude~\cite{bob,musolf94}. Such experiments involve the
scattering of longitudinally polarized electrons from unpolarized
targets, allowing the determination of a parity-violating asymmetry
$\APV\equiv{(\sigma_R-\sigma_L)}/{(\sigma_R+\sigma_L)}$, where
$\sigma_{R(L)}$ is the cross section for incident right(left)-handed
electrons.  $\APV$ arises from the interference of the weak and
electromagnetic amplitudes \cite{zeld}. 
Typical asymmetries are small, ranging from
0.1 to 100 parts per million (ppm).

Four experiments have published $\APV$ measurements in elastic
electron-proton scattering. 
The SAMPLE result~\cite{sample} 
at backward angle constrained $G_M^s$
at $Q^2\sim 0.1 \gevc$. 
The HAPPEX~\cite{happex}, A4~\cite{A41,A42}, and G0~\cite{G0} results 
at forward angle
constrained a linear combination of $G_E^s$ and $G_M^s$
in the range $0.1 < Q^2 < 1 \gevc$.
While no measurement independently indicates a significant
strange form factor contribution, the A4 measurement
at $Q^2=0.108 \gevc$ and the G0 measurement at slightly higher $\qsq$
each suggest a positive deviation, at the level of $\sim 2\sigma$, 
from the asymmetry which would be expected with no strange quark 
contribution~\cite{A42,G0}.

In this paper, we report a new measurement of 
$\APV$ in elastic electron-proton 
scattering at $Q^2\sim 0.1 \gevc$. This first result from experiment 
E99-115 at the Thomas Jefferson National Accelerator Facility (JLab) has 
achieved the best precision on $\APV$ in electron-nucleon 
scattering.
The sensitivity of the measurement to strange form factors is 
similar to that of the recently published A4 measurement~\cite{A42}.
However, while A4 counted individual electron scattering events,
the measurement reported here employs an analog integrating technique,
described below,  
that will ultimately allow for very small overall statistical and 
systematic uncertainty. 
The ultimate goal of the experiment is to reach a 
precision \mbox{$\delta(\APV)\sim 0.1$ ppm}.

\section{Description of the Apparatus}

The experiment is situated in Hall A at JLab. A \mbox{$35~\mu$A} 
continuous-wave beam of longitudinally polarized 3.03~GeV electrons
is incident on a 20~cm long liquid hydrogen target. 
The highly polarized (75-85\%) electron beam is generated from a 
strained-layer GaAs photocathode using circularly polarized 
laser light.
Scattered electrons are focused by twin spectrometers onto 
total-absorption detectors situated in heavily-shielded detector
huts, creating a clean separation between elastically scattered
electrons and inelastic backgrounds. The spectrometers are
arranged to create an approximately left-right symmetric acceptance.

Two separate detector segments in each 
spectrometer arm cover the full flux of elastically scattered 
electrons, for a total of 4 detector photomultiplier tubes (PMTs). 
The PMT response is integrated; the detector elements and the 
associated electronics are designed to accept an elastic flux 
rate of $\sim100$~MHz at full design luminosity.

The experimental configuration is similar to the previous measurement of 
$\APV$ at $Q^2\sim 0.5$ $\gevc$~\cite{happex}. The presently reported
measurement is enabled by the addition of septum magnets to accept
very-forward scattered electrons with 
$\langle\theta_{lab}\rangle\sim 6^{\circ}$, and the introduction
of radiation-hard focal plane detectors which can survive the increased
scattered electron rate. This configuration is described in more 
detail in the recent report on the $\APV$ result with a $^4$He 
target~\cite{helium4}.

The helicity of the polarized
electron beam is set every 33.3 ms; each of these periods
of constant helicity will be referred to as a ``window.''  The helicity
sequence is structured as pairs of windows with opposite helicity 
(``window pairs''), with the helicity of the first window selected 
pseudo-randomly.  The integrated response of the detector PMTs, 
beam current monitors, and beam position monitors is digitized and 
recorded into the data stream for each window.

\section{Data Sample and Analysis}

The data sample consists of roughly 11 million helicity-window pairs.
Loose requirements are imposed on beam quality 
which remove periods of current, position, or energy
instability from the final data set.
However, no helicity-dependent cuts are applied.
After applying selection criteria, 
$9.96\times10^6$ window pairs remain for further analysis. 

The right-left asymmetry in the integrated detector response,
normalized to the average beam current for each window,
is computed for each window pair and then corrected
for fluctuations in the beam trajectory to form the raw asymmetry
$\ARAW$. The first order dependence on five correlated beam 
parameters (energy and horizontal and vertical position and angle) is removed
by two independent analysis methods; the numerical difference between the 
two results is negligible compared to the final statistical uncertainty.
\begin{figure}
\includegraphics[width=0.99\columnwidth]{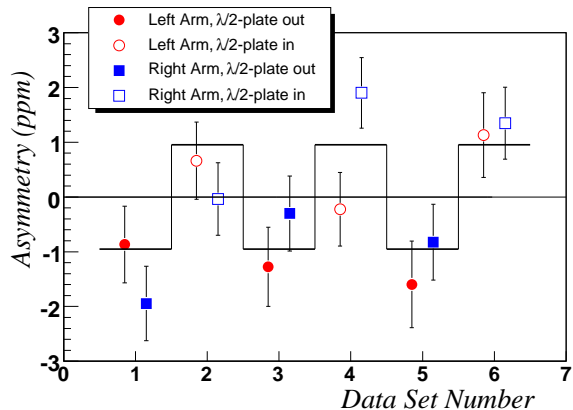}
\caption{$\ARAW$ for all data, grouped by 
$\lambda$/2-plate state in sequential samples. The circles and squares
represent the average of the 2 PMT channels in each spectrometer arm, and
the line represents $\ARAW$, averaged over the run and plotted
with the appropriate sign for each half-wave plate state.}
\label{figure:slugplot}
\end{figure}

The $\ARAW$ window-pair distribution has an RMS of $\sim 620$ ppm.
Non-Gaussian tails are negligible over more than 4 orders of magnitude. This
demonstrates that the
distribution is dominated by the counting statistics of an 
elastically scattered electron rate of $\sim 40$ MHz. Contributions to the 
fluctuations from background, electron beam, electronic noise or target 
density are negligible.

The cumulative correction for $\ARAW$ due to helicity-correlated 
differences in electron beam parameters is $-0.079\pm 0.032$ ppm. This
correction is small compared to the statistical error on $\ARAW$ due to 
several important factors. First, careful attention is given to the design and
configuration of the laser optics in the polarized source
to reduce helicity-correlated beam asymmetries to a manageable level.
Over the duration of data collection, the cumulative helicity-correlated
asymmetries in the electron beam are 0.022 ppm in energy, 8 nm in 
position, and 4 nrad in angle. 

Additionally, the asymmetry averaged 
over all PMTs is quite insensitive to the beam trajectory due to the 
symmetric detector configuration. 
The largest correction of -0.130 ppm is from the beam monitor that 
is predominantly sensitive to the helicity-correlated beam energy asymmetry. 
The systematic error in the correction 
is estimated by studying residual correlations of
beam asymmetries with the responses of individual PMTs, which
are significantly more sensitive to the beam trajectory 
due to the division of the elastic peak over the detector segmentation.

The effect of charge normalization is a 2.6~ppm correction
to the detector-response asymmetry.
Dedicated calibration runs are used to constrain the relative alinearity
between the beam monitors and the detectors ($<0.2$\%) and the absolute
alinearity of the detector PMTs ($<1$\%).
No alinearity correction to $\ARAW$ is applied, while 
an uncertainty of 0.015~ppm is assigned.

A half-wave ($\lambda$/2) plate is periodically inserted into the 
laser optical path, passively reversing the
sign of the electron beam polarization. Roughly equal statistics are thus 
accumulated with opposite signs for the measured asymmetry, which suppresses 
many systematic effects. Figure~\ref{figure:slugplot} shows 
$\ARAW$ for all data, averaged over the 2 PMT channels in each spectrometer,
grouped by $\lambda$/2-plate state and divided 
into 6 sequential samples. 
The observed fluctuations are consistent with purely statistical
fluctuations around the average parity-violating asymmetry, shown on the plot
with the expected sign flip due to half wave plate state, with a 
$\chi^{2}$ per degree of freedom of 1.0.

The physics asymmetry $\APV$ is formed from $\ARAW$ 
by correcting for beam polarization, backgrounds, and finite acceptance:
\begin{equation}
\APV = \frac{K}{P_b}\frac{\ARAW - P_b\sum_{i} A_{i}f_{i}}{1-\sum_{i} f_{i}}
\end{equation}
where $P_b$ is the beam polarization, $f_{i}$ are background fractions and
$A_{i}$ the associated background asymmetries, and $K$ accounts for the range
of kinematic acceptance.

The beam polarization measured by the Hall A Compton
polarimeter~\cite{HallA_Compton} is determined to be 
$P_{b} = 0.813 \pm 0.016$, averaged over the duration of the run.
The result is consistent, within systematic uncertainties, 
with dedicated polarization measurements
using M{\o}ller scattering in Hall A and Mott scattering in
the low-energy injector.

The average $Q^{2}$ is determined to be 
$\langle Q^{2} \rangle = 0.099 \pm 0.001 \gevc$ by dedicated
low-current runs; the uncertainty in this value contributes to
the systematic error of the asymmetry.
The acceptance correction to account for the non-linear dependence of the 
asymmetry with $\qsq$ is computed, using a Monte Carlo simulation, to be 
$K=0.976\pm 0.006$.

Largely due to the excellent hardware resolution of the spectrometers 
($\delta p/p < 0.1$\%), the total dilution
to the PMT response 
from all background sources is less than 1\%. The largest contribution 
of 0.9\%\ comes from the aluminum windows of the cryogenic target. The 
asymmetry of the background is of the same sign and similar magnitude to 
that of $\APV$ from elastic scattering off hydrogen, which reduces its 
effect on the measurement.

While inelastic scattering backgrounds do not directly reach the detectors, 
dedicated runs are used to estimate the contribution 
from charged particles which rescatter inside the spectrometers. Rates
in the detectors are studied as the central 
spectrometer momentum is varied. 
Individual scattered electrons are tracked, using
drift chambers at low beam currents, to determine the location of rescattering
in the spectrometer. 
From these studies, an upper limit
on $\APV$ due to possible rescattering from polarized iron or 
unpolarized material is determined to be 0.031 ppm. 

The corrections are summarized in Table~\ref{table:Acorrections}.
After all corrections, the result at $\qsq = 0.099 \gevc$ is
\begin{equation}
\APV = -1.14 \pm 0.24 \,\,\mbox{(stat)} \,\pm 0.06 \,\,\mbox{(syst) ppm.}
\end{equation}
Additional details of this analysis are given in~\cite{helium4}.

\begin{table}
\begin{center}
\vspace{0.25cm}\begin{tabular}{lrcl}
\multicolumn{4}{l}{Correction (ppm)}\\ \hline \hline
Target windows & $0.006$  & $\pm$ & $0.016$ \\ 
Rescatter &      $0.000$  & $\pm$ & $0.031$ \\ 
Beam Asyms. &    $-0.079$ & $\pm$ & $0.032$  \\ 
Alinearity &     $0.000$   & $\pm$ & $0.015$  \\ 
\hline \hline \\
\multicolumn{4}{l}{Normalization Factors}\\ \hline \hline
Polarization $P_b$ & $0.813$ & $\pm$ & $0.016$ \\ 
Acceptance $K$     & $0.976$ & $\pm$ & $0.006$ \\ 
$Q^{2}$ Scale      & $1.000$  & $\pm$ & $0.015$ \\ \hline \hline
\end{tabular}
\caption{Corrections to $\ARAW$ and systematic uncertainties.
\label{table:Acorrections}}
\end{center}
\end{table}

\section{Results and Conclusions}

This parity-violating asymmetry is given in 
the standard model by:
\begin{eqnarray}
\APV &=& - \frac{G_{F}Q^{2}}{4\pi\alpha\sqrt{2}}
\times \left\{(1+R_V^p)(1-4\sin^{2}\theta_{W})\right.\nonumber \\
&&-\left.(1+R_V^n)\frac{\epsilon G^{\gamma p}_{E} G^{\gamma n}_{E} + \tau G^{\gamma p}_{M} G^{\gamma n}_{M}}
{\epsilon(G^{\gamma p}_{E})^{2} + \tau(G^{\gamma p}_{M})^{2}}\right.\nonumber \\
&&-\left.(1-R_V^{(0)})\frac{\epsilon G^{\gamma p}_{E} G^{s}_{E} + \tau G^{\gamma p}_{M} G^{s}_{M}}
{\epsilon(G^{\gamma p}_{E})^{2} + \tau(G^{\gamma p}_{M})^{2}}\right.\nonumber \\
&&-\left.\frac{(1-4\sin^{2}\theta_{W})\,\epsilon^{\prime}\,G^{\gamma p}_{M}}
{\epsilon(G^{\gamma p}_{E})^{2} + \tau(G^{\gamma p}_{M})^{2}}\left[-2\,(1+R_A^{T=1})\,G^{T=1}_{A}\right.\right.\nonumber \\
&&+\left.\left.(\sqrt{3}R_A^{T=0})\,G^{T=0}_{A}\right]\right\}
\end{eqnarray}
where $G^{\gamma p(n)}_{E(M)}$ are the proton (neutron) 
electric (magnetic) form-factors,
$G^{T=1(0)}_{A}$ is the isovector (isoscalar) proton axial form factor,
$G_{F}$ is the Fermi constant,
$\alpha$ is the fine structure constant, and $\theta_{W}$ is the
electroweak mixing  angle.  All form factors are functions of $Q^{2}$,
and $\epsilon=0.994$, $\tau=0.028$, $\epsilon^{\prime}=0.018$ are 
kinematic quantities.
The $R_{V,A}$ factors parametrize the electroweak radiative corrections
of the neutral weak current~\cite{musolf94}. All the vector 
corrections~\cite{musolf94} and the axial corrections~\cite{zhu2000}
are converted to their ($\overline{MS}$) values with
$\sin^{2}\theta_{W}\equiv\sin^{2}\hat{\theta}_{W}(M_Z)=0.23120(15)$~\cite{PDG}.
Corrections due to purely electromagnetic radiative corrections 
are negligible due to the small momentum acceptance ($\delta p/p<3\%$) and 
the spin independence of soft photon emission~\cite{maximon}.

The values for the electromagnetic form factors $G^{\gamma p(n)}_{E(M)}$
are taken from a recently published phenomenological fit to world data
at low $Q^{2}$~\cite{walcher_friedrich}, with uncertainties in each value
based on error bars of data near $Q^{2}=0.1 \gevc$.
The values (and
relative uncertainty) used are: $G^{p}_{E} = 0.754$ (2.5\%),  
$G^{p}_{M} = 2.144$ (1.5\%), $G^{n}_{E} = 0.035$ (30.0\%), and  
$G^{n}_{M} = -1.447$ (1.5\%).
The contribution from axial form factors 
is calculated to be $0.026 \pm 0.008$~ppm at these kinematics.

At the central kinematics, $\APV$ is estimated ($G^s=0$) to be
$\APV^{(s=0)} = -1.43\, \pm \,0.11 \,\mbox{(FF) ppm}$ where the error
comes mainly from the uncertainty in $G^{n}_{E}$.
We thus extract a
measurement of the combination of strange form-factors:
$G^{s}_{E} + 0.080 \, G^{s}_{M} = 0.030 \, \pm \, 0.025 \, 
\mbox{(stat)} \, \pm \, 0.006 \, \mbox{(syst)} \, \pm \, 0.012 \, 
\mbox{(FF)}$ at \mbox{$Q^{2}=0.099 \gevc$}.

\begin{figure}
\includegraphics[width=1.0\columnwidth]{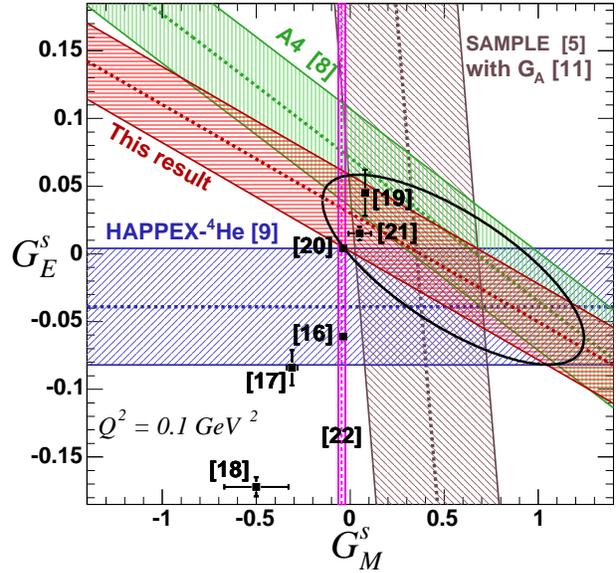}
\caption{The four $\APV$ measurements at 
$Q^2 = 0.09$-$0.11\gevc$ are shown, with shaded bands representing the 
1-sigma combined statistical and systematic uncertainty. 
Also shown is the combined 95\%\ C.L. ellipse
from all four measurements. The black squares and narrow vertical band
represent various theoretical calculations (\cite{park}-\cite{leinweber}).}
\label{figure:gegm}
\end{figure}

This result is displayed in Figure~\ref{figure:gegm}, 
along with three other published strange form factor
measurements. Each of these measurements was carried out in a narrow $\qsq$
range of 0.09-0.11$\gevc$ such that combining them introduces no
significant additional uncertainty. From the four measurements
shown in the figure, limits on $\ges$ and $\gms$ at $\qsq \sim0.1\gevc$ are
extracted without any additional assumptions.
The absence of theoretical guidance
for the $\qsq$ dependence of the form factors precludes the use of
published data from higher $\qsq$ for this fit.  The 95\% allowed
contour from the combined fit is shown in Figure~\ref{figure:gegm}. 
The best fit values are $G_E^s = -0.01 \pm 0.03$ 
and $G_M^s = +0.55 \pm 0.28$.
While this fit favors positive values 
for $G_M^s$, the origin ($G^s=0$) is still allowed at the 95\%~C.L. \,
Figure~\ref{figure:gegm} also shows results from various theoretical
calculations \cite{park}-\cite{leinweber}.

In conclusion, we report a precise measurement of $\APV$ in elastic
electron-proton scattering at $Q^2=0.099 \gevc$ which has resulted in
improved constraints on the strange form factors at $Q^2\sim 0.1 \gevc$.
The HAPPEX measurements at $Q^2\sim 0.1 \gevc$ from both $^1$H 
and $^4$He targets will be improved by a factor of 2 to 3 in precision
by additional data collected in late 2005. Given the currently allowed region
in Fig.~\ref{figure:gegm}, such precision has the potential to 
dramatically impact our understanding of the role of strange 
quarks in the nucleon.

\begin{ack}
We wish to thank the entire staff of JLab for their efforts to develop and
maintain the polarized beam and the experimental apparatus. 
This work was supported by DOE contract DE-AC05-84ER40150 Modification No.~M175, 
under which the Southeastern Universities Research Association (SURA) 
operates JLab, and by the Department of Energy, the National 
Science Foundation, the INFN (Italy), 
the Natural Sciences and Engineering Research Council of Canada, and the 
Commissariat \`a l'\'Energie Atomique (France).
\end{ack}


\begin{thebibliography}{99}
\bibitem{aneesh}
D.~B.~Kaplan and A.~Manohar, Nucl.\ Phys.\ B {\bf 310} (1988) 527.
\bibitem{bob}
R.~D.~McKeown, Phys.\ Lett.\ B\ {\bf 219} (1989) 140.
\bibitem{musolf94}
M.~J.~Musolf {\it et al.}, 
Phys.\ Rep.\ {\bf 239} (1994) 1.
\bibitem{zeld}
Ya. B. Zel'dovich, Sov. Phys. JETP, {\bf 36} (1959) 964.
\bibitem{sample}
D.~T.~Spayde {\it et al.}, Phys.\ Lett.\ B\ {\bf 583} (2004) 79.
\bibitem{happex}
K.~A.~Aniol {\it et al.}, Phys.\ Lett. B\ {\bf 509} (2001) 211; K.~A.~Aniol {\it et. al.}, Phys.\ Rev.\ C {\bf 69} (2004) 065501.
\bibitem{A41}
F.~E.~Maas {\it et al.}, Phys.\ Rev.\ Lett.\ {\bf 93} (2004) 022002.
\bibitem{A42}
F.~E.~Maas {\it et al.}, Phys.\ Rev.\ Lett.\  {\bf 94} (2005) 152001.
\bibitem{G0}
D.~S.~Armstrong {\it et al.}, Phys. Rev. Lett. {\bf 95} (2005) 092001.
\bibitem{helium4}
K.~A.~Aniol {\it et al.}, nucl-ex/0506010, accepted for publication in Phys. Rev. Lett.
\bibitem{HallA_Compton}
S.~Escoffier {\it et al.}, physics/0504195.
\bibitem{zhu2000}
S.-L.~Zhu {\it et al.}, 
Phys.\ Rev.\ D\ {\bf 62} (2000) 033008.
\bibitem{PDG}
J.~Erler and P.~Langacker, Phys. Lett. B {\bf 592} (2004) 114.
\bibitem{maximon}
L.~C.~Maximon and W.~C.~Parke, Phys.\ Rev.\ C\ {\bf 61} (2000) 045502.
\bibitem{walcher_friedrich}
J.~Friedrich and Th.~Walcher, Eur.\ Phys.\ J.\ A\ {\bf 17} (2003) 607.
\bibitem{park}
N.~W.~Park and H.~Weigel, Nucl.\ Phys.\ A {\bf 451} (1992) 453. 
\bibitem{drechsel}
H.~W.~Hammer, U.~G.~Meissner, and D.~Drechsel, Phys.\ Lett.\ B\ {\bf 367} (1996) 323.
\bibitem{hammer}
H.-W.~Hammer and M.~J.~Ramsey-Musolf, Phys.\ Rev.\ C\ {\bf 60} (1999) 045204.
\bibitem{silva}
A.~Silva {\it et al.}, Phys.\ Rev.\ D\ {\bf 65} (2001) 014015.
\bibitem{lyub}
V.~Lyubovitskij {\it et al.}, Phys.\ Rev.\ C\ {\bf 66} (2002) 055204.
\bibitem{lewis}
R.~Lewis {\it et al.}, Phys.\ Rev.\ D\ {\bf 67} (2003) 013003.
\bibitem{leinweber}
D.~B.~Leinweber {\it et al.}, Phys. Rev. Lett. {\bf 94} (2005) 212001.
\end{thebibliography}
\end{document}